\documentclass[twocolumn]{aastex}
\usepackage{amsmath}
\usepackage{xcolor}

\usepackage{booktabs}

\newcommand{\del}[1]{\textcolor{red}{{\iffalse{#1}\fi}}}
\usepackage{hyperref}

\begin{document}

\title{ Formation of Binary Millisecond Pulsars with Helium White Dwarfs in a New Magnetic Braking Prescription}

\author[0009-0001-6764-106X]{Xing-Peng Yang}
\affil{School of Science, Qingdao University of Technology, Qingdao 266525, People’s Republic of China;chenwc@pku.edu.cn}
\author[0000-0002-0785-5349]{Wen-Cong Chen}
\affil{School of Science, Qingdao University of Technology, Qingdao 266525, People’s Republic of China;chenwc@pku.edu.cn}
\affil{School of Physics and Electrical Information, Shangqiu Normal University, Shangqiu 476000, People’s Republic of China}

\begin{abstract}
Magnetic braking (MB) mechanism plays a vital role throughout the evolution of low-mass X-ray binaries (LMXBs). Considering the standard MB prescription, the initial orbital periods of LMXBs that can evolve into binary millisecond pulsar (MSP) with He white dwarfs (WDs) and short orbital periods ($2-9~\rm hours$) are within an extremely narrow interval, which was named the fine-tuning problem. Employing the detailed binary evolution model, we investigate the evolution of LMXBs in both the standard and convection and rotation boosted (CARB) MB laws. In the standard MB case, it is difficult for donor stars to form a He core and exhaust H envelope through mass transfer at short orbital periods, making them semidetached systems. The CARB MB mechanism can drive LMXBs evolve toward compact detached MSP-WD systems in wide initial orbital periods, over which binary MSPs with long orbital periods will be produced. We obtain the initial parameter space of binary MSPs with He WDs in the initial orbital period and donor-star mass plane, which can be applied to future statistics study by population synthesis simulations. We also discuss a new relation between orbital period and WD mass, formation of persistent ultra-compact X-ray binaries with relatively long orbital periods, and detectability of compact MSP-WD systems as low-frequency gravitational wave sources.
\end{abstract}

\section{Introduction}
Millisecond pulsars (MSPs) are rapidly rotating (spin periods are less than 30 ms) neutron stars (NSs) with very small period derivatives ($\sim10^{-21}-10^{-19}~\rm s\,s^{-1}$). Because of extremely stable periods and very small spin-down rates, MSPs are thought to be the most precise celestial clocks \citep{bhat22}. Assuming that radio MSPs consume their rotation energy via magnetic dipole radiation, it is inferred that their magnetic fields are in the range of $10^8-10^9~\rm G$. It is generally thought that MSPs were spun up to millisecond periods by accreting matter and angular momentum from their low-mass ($<1-2~M_\odot$) companion stars in low-mass X-ray binaries \citep[LMXBs,][]{alpa82,bhat91}. The discovery of accreting MSPs provided strong support for the link between binary MSPs and LMXBs \citep{wijn98,arch09}. 

In binary MSPs, companion stars include degenerate stars and non-degenerate stars. Most MSPs were discovered to possess a Helium (He) white dwarf (WD) or a Carbon–Oxygen WD companion. However, binary MSPs with short orbital periods ($P_{\rm orb}<1~\rm day$) also include eclipsing pulsars either with a non-degenerate main-sequence (MS) companion with a mass in the range $0.1–0.8~M_\odot$ (redbacks) or with a $0.01–0.05~M_\odot$ brown dwarf \citep[black widows,][]{taur23}. 

The progenitors of binary MSPs with relatively long orbital periods ($>1~\rm day$) are LMXBs with wide orbits, in which the donor star fills the Roche lobe and begins a Case B mass transfer when it had evolved into a (sub)giant \citep{rapp95,taur99,pods02}. There exist several known binary MSPs with He WD companions (the masses are in the range of $0.13-0.21~M_\odot$) in very narrow orbits \citep[orbital periods are shorter than 9 hours,][]{istr14}. These systems should evolve from LMXBs with tight orbits where the donor star begins a Case A mass transfer on the MS or Hertzsprung gap stage. These compact MSP-WD systems will coalesce and evolve into Ultracompact X-ray binaries (UCXBs) within a Hubble time, and appear as low-frequency gravitational wave (GW) sources that can be detected by LISA slightly before and during the UCXB stage \citep{taur18,chen20}. Binary MSPs can be used to test general relativity and the equation of state of NSs, as well as to detect nanohertz GWs by MSP timing arrays \citep{taur23}. Especially, they are also evolutionary fossils of LMXBs, becoming ideal probes for testing stellar and binary evolutionary theory. Therefore, there is a broad astrophysical significance for studying the formation and evolution of binary MSPs.

To reproduce the known compact MSP-WD systems with orbital periods of $2-9~\rm hours$, \cite{istr14} had performed a detailed stellar evolution model for the evolution of $\sim400$ NS-MS binaries based on the standard MB prescription given by \cite{Rapp83}. When the initial mass of the MS donor star is $1.4~M_\odot$, a severe fine-tuning is inevitable for the initial orbital period of those NS-MS binaries in order to evolve toward compact MSP-WD systems even if they take an anomalously large MB index of $\gamma=5$ \citep{istr14}. Similar results were also noticed in some previous works on the evolution of LMXBs \citep{pods02,lin11,jia14} and formation of binary MSP PSR J0348+0432 \citep{anto13}. \cite{rome19} proposed that the fine-tuning problem in the formation of compact MSP-WD systems can be alleviated by adopting a stronger MB prescription given by \cite{van19mn}. \cite{chen21} confirmed that the fine-tuning problem can be solved with a relatively efficient intermediate MB prescription given by \cite{van19mn}.  However, this MB prescription cannot produce the known wide-orbit binary MSPs \citep{chen21}. Therefore, further research on the formation of binary MSPs is needed for some alternative MB prescriptions.

\cite{Van19} proposed a convection and rotation boosted (CARB) MB prescription to reproduce the detected mass-transfer
rates, orbital periods, and mass ratio of all known persistent Galactic NS LMXBs. The observed characteristics of binary pulsars were also found to be reproduced by the CARB MB law \citep{deng21}. Recently, \cite{wei24} used the binary MSP PSR J1012+5307 with an extremely low-mass WD to test different MB models and confirm that intermediate MB and CARB MB laws can successfully reproduce its observed parameters except for the effective temperature of the WD.

In this work, we attempt to investigate the origin of fine-tuning problem in forming compact MSP-WD systems. We also use the CARB MB law to model the formation and evolution of MSP-WD systems and obtain the initial parameter space of their progenitors. In Section 2, we introduce the binary evolution code and the MB laws. Section 3 gives our detailed results by binary evolution model under the standard and CARB MB laws. Finally, we present a brief discussion and summary in Sections 4 and 5, respectively. 

\section{Binary evolution code and the CARB MB law}
\subsection{Binary Evolution Code}
To investigate the fine-tuning problem on the formation of compact binary MSPs consisting of a MSP and a He WD, we employ the binary module in the Modules for Experiments in Stellar Astrophysics \citep[MESA version r12115;][]{paxt11,paxt13,Paxton15,paxt18,paxt19} to simulate the evolution of binary systems consisting of a NS (with a mass of $M_{\rm NS}$) and a MS companion star (with a mass of $M_{\rm d}$) in a circular orbit. The NS is considered to be a point mass, and the code only models the orbital evolution and nuclear synthesis of the MS star with a solar composition (i.e. $X = 0.7, Y = 0.28, Z = 0.02$). We run the detailed MESA model for each system until the stellar age exceeds the Hubble time (14 Gyr) or the time step reaches a minimum time step limit.

When the MS star fills the Roche lobe, the mass transfer initiates from the donor star to the NS at a rate $\dot{M}_{\rm tr}$. Assuming the accretion efficiency is $f$, the accretion rate of the NS is $\dot{M}_{\rm acc}=f\dot{M}_{\rm tr}$. In this work, we take $f=1$. During the accretion, it is generally thought that the mass-growth rate of the NS is $\dot{M}_{\rm NS}={\rm min}(\dot{M}_{\rm Edd},f\dot{M}_{\rm tr})$, which is limited by the Eddington accretion rate \citep{Tauris06}
\begin{equation}
    \dot{M}_{\rm Edd}=1.5\times10^{-8}~M_{\odot}\,\rm yr^{-1}.
\end{equation}
The excess material in unit time ($\dot{M}_{\rm acc}-\dot{M}_{\rm NS}$) is assumed to be ejected from the vicinity of the NS, carrying away the specific orbital angular momentum of the NS. 

During and after He burning, we use Type 2 opacities to simulate extra C/O burning. For the ratio between the mixing length and the height of the local pressure scale, we adopt $\alpha=2$. Once the mass transfer starts, the "Ritter" mass transfer scheme is arranged \citep{ritt88}. In the CARB MB prescription, it requires a loss rate of stellar winds, which is an important input parameter in the rate of angular momentum loss (see also Section 2.2). The "Dutch" wind with a scaling factor of 1.0 is chosen in the wind setting option schemes that include $hot_{-}wind_{-}scheme$ \citep{gleb09}. 

\subsection{Influence factors of orbital evolution}
The total orbital angular momentum of a NS-MS binary is $J=\Omega a^2M_{\rm d}M_{\rm NS}/(M_{\rm d}+M_{\rm NS})$, where $\Omega$ and $a$ are the orbital angular velocity and separation of the binary, respectively. Combining Kepler's third law $a^3=G(M_{\rm d}+M_{\rm NS})/\Omega^2$ ($G$ is the gravitational constant), the orbital period derivative of the system obeys the follow expression as
\begin{equation}
    \frac{\dot{P}_{\rm orb}}{P_{\rm orb}}=3\frac{\dot{J}}{J}-3\frac{\dot{M}_{\rm d}}{M_{\rm d}}\left[1 - q\eta - \frac{q(1 - \eta)}{3(1 + q)}\right]
\end{equation}
where $q=M_{\rm d}/M_{\rm NS}$ is the mass ratio of the system, $\eta=-\dot{M}_{\rm NS}/\dot{M}_{\rm d}$ is the accretion efficiency of the NS.

During the evolution of NS binaries, the total rate of orbital angular momentum loss is given by
\begin{equation}
\dot{J}=\dot{J}_{\rm gr}+\dot{J}_{\rm mb}+\dot{J}_{\rm ml},   
\end{equation}
where $\dot{J}_{\rm gr}, \dot{J}_{\rm mb}$, and $\dot{J}_{\rm ml}$ are the rates of orbital angular momentum loss caused by gravitational radiation (GR), MB, and mass loss, respectively. During the accretion, the rate of orbital angular momentum loss due to mass loss can be written as
\begin{equation}
\dot{J}_{\rm ml} = 
\begin{cases}
 -\frac{\Omega a^2M_{\rm d}^2(\dot{M}_{\rm tr} -\dot{M}_{\rm Edd})}{(M_{\rm NS}+M_{\rm d})^{2}}, & \dot{M}_{\rm acc}>\dot{M}_{\rm Edd} \\
 0, & \dot{M}_{\rm acc} \le \dot{M}_{\rm Edd}.
\end{cases}
\end{equation}
It is clear that angular momentum loss would cause orbital decay according to the first term on the right-hand side of Equation (2). Because $\eta<1$ and $q<1$ for a low-mass donor star ($M_{\rm d}<1.4~M_\odot$), the mass transferred from the less massive donor star to the more massive NS could lead to orbital expansion. It depends on the competition between these two effects for the orbital evolution fates of NS-MS binaries.

\subsection{Two MB laws}
\subsubsection{Standard MB prescription}
For those low-mass stars with masses $\le 1.5~\rm M_{\odot}$, the existence of convective envelope causes an efficient spin angular momentum loss via magnetic winds \citep{Verbunt81}. However, the tidal interaction between the two components would spin the star up to corotate with the orbital motion, so magnetic winds indirectly extract orbital angular momentum \citep{patt84}. In the standard MB prescription given by \cite{Rapp83}, the rate of angular momentum is
\begin{equation}
\dot{J}_\mathrm{mb} = -6.82\times10^{34}\left(\frac{M_{\rm d}}{M_\odot}\right)\left(\frac{R_{\rm d}}{R_\odot}\right)^{\gamma}\left(\frac{P_{\rm orb}}{1~\rm day}\right)^{-3} \rm g\,cm^2s^{-2},
\end{equation}
where $R_{\rm d}$ and $P_{\rm orb}$ are the radius of the donor star and orbital period of the system, respectively. In this work, we take the magnetic braking index as $\gamma=4.0$.

\subsubsection{CARB MB prescription}

In a modified MB law, the stellar winds' velocity was revised from the escapse velocity $v_{\rm esc}$ to $(v_{\rm esc}^{2}+2\Omega^{2}R_{\rm d}^{2}/K^{2})^{1/2}$ \citep{Matt12}, where $K = 0.07$ is a constant fitted by a grid of simulations \citep{revi15}. Furthermore, the influence of stellar rotation and convective eddy turnover timescale on the surface magnetic field was also included \citep{park71,noye84,ivan06}. Therefore, the rate of angular momentum loss given by the proposed convection and rotation boosted (CARB) MB prescription is \citep{Van19}
\begin{align}
\dot{J}_{\rm mb}=-\frac{2}{3}\dot{M}_{\rm w}^{-1/3}R_{\rm d}^{14/3}(v_{\rm esc}^2+2\Omega^2R_{\rm d}^2/K^2)^{-2/3}\nonumber\\ \times\Omega_{\odot}B_{\odot}^{8/3}(\frac{\Omega}{\Omega_{\odot}})^{11/3}(\frac{\tau_{\rm conv}}{\tau_{\odot,\rm conv}})^{8/3},
\end{align}
where $\dot{M}_{\rm w}$ is the loss rate of stellar winds, $\tau_{\rm conv}$ is the turnover time of convective eddies of the donor star, $\Omega_\odot \approx 3 \times 10^{-6} ~\rm s^{-1}$, $B_\odot = 1.0~ \rm G$, and $\tau_{\odot, \rm conv} = 2.8 \times 10^6 ~\rm s$ are the Sun's surface rotation rate, surface magnetic field strength, and convective turnover time, respectively. If the donor star has both a convective envelope and a radiative core, the standard MB or CARB MB law is arranged to operate in the MESA code.

\begin{figure}
\centering
\includegraphics[width=1.0\linewidth,trim={0 0 0 0},clip]{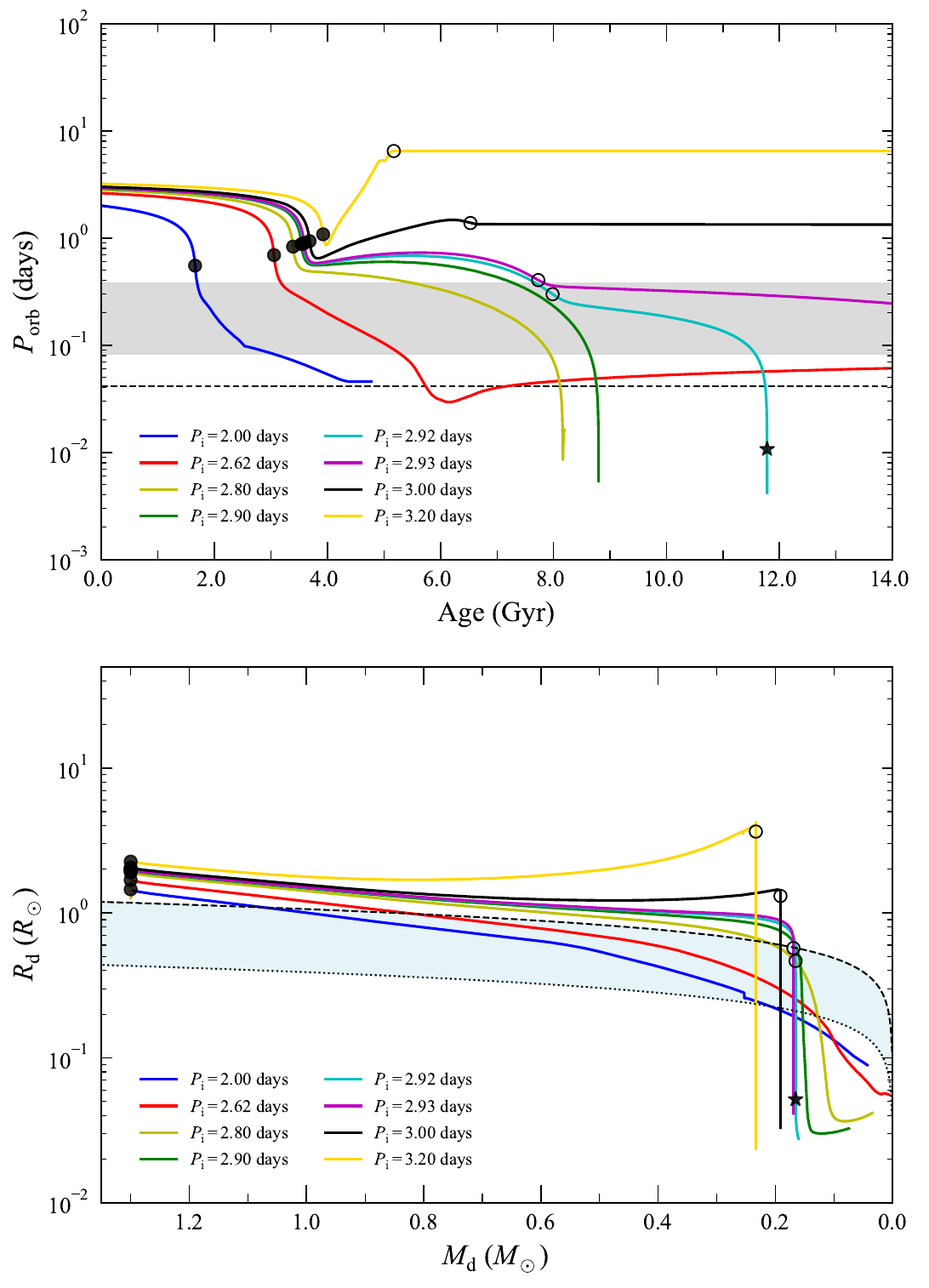}
\caption{Evolution of NS-MS binaries with $M_{\rm NS,i}=1.4~M_{\odot}$, $M_{\rm d,i}=1.3~M_{\odot}$, and $P_{\rm orb,i}=1.6-3.2~\rm days$ in the orbital period vs. stellar age diagram (top panel) and donor-star radius vs. donor-star mass diagram (bottom panel) under the standard MB law. The solid circles, open circles, and solid stars denote the onset of the first mass transfer, the end of the first mass transfer, and the onset of the UCXB stage, respectively. In the top panel, the horizontal dashed line and shaded area represent an orbital period of $60~\rm minutes$ and an orbital period range of $2-9~\rm hours$, respectively. In the bottom panel, the black dashed and dotted curves describe the evolutionary tracks of Roche lobe radii of binary systems with a $1.5~M_\odot$ NS in orbits with orbital periods of 9 and 2 hours, respectively.}
\label{s1}
\end{figure}

\section{Simulated Results}

To investigate the origin of the fine-tuning problem, we first model the evolution of NS-MS systems consisting of a NS with an initial mass of $M_{\rm NS,i}=1.40~M_{\odot}$ and a MS donor star with an initial mass of $M_{\rm d,i}=1.30~M_{\odot}$, which are in circular orbits with initial orbital periods of $P_{\rm orb,i}=2.0-3.2~\rm days$ and $P_{\rm orb,i}=1.3-25.0~\rm days$ for the standard MB and CARB MB cases, respectively.

\subsection{Origin of the fine-tuning problem}
Considering the standard MB prescription, Figure \ref{s1} plots the evolutionary tracks of NS-MS binaries in the orbital period versus the stellar age plane (top panel) and in the donor-star radius versus the donor-star mass plane (bottom panel). Once donor stars fill their Roche lobes, the mass transfer proceeds and the systems appear as LMXBs. During the evolution of LMXBs, NSs are spun up to millisecond periods by accreting matter and angular momentum from low-mass donor stars. Donor stars only in two NS-MS binaries with $P_{\rm orb,i}=2.92$ and $2.93~\rm days$ can decouple from their Roche lobes at short orbital periods of $2-9~\rm hours$. During the first mass transfer, donor stars in these two systems are removed most of the H envelope and remain He cores. Subsequently, the He cores evolved into He WDs through the contraction and cooling phases. Therefore, these two systems can evolve into detached MSP-WD systems with short orbital periods of $2-9~\rm hours$ after the first mass transfer stops. Those systems with $P_{\rm orb,i}>2.93~\rm days$ would evolve towards MSP-WD systems with orbital periods longer than $0.7~\rm days$. The remaining systems (with initial orbital periods shorter than 2.90 days) always proceed with a stable mass transfer and cannot evolve into detached systems. Therefore, the progenitors of binary MSPs with He WDs and short orbital periods ($2-9~\rm hours$) have an ultra-narrow initial orbital period range, which was named the fine-tuning problem of detached MSP-WD systems with short orbital periods \citep{istr14}.

To evolve toward detached MSP-WD systems with short orbital periods, donor stars have to decouple from the Roche lobes when orbital periods decrease to a range of 0.2 to 0.4 days. Meanwhile, donor stars must exhaust most of their hydrogen envelope via stable mass transfer to remain He cores. According to the relation between orbital periods and WD masses in binary MSPs \citep{jia14}, WD masses concentrate on a narrow range of $0.16-0.17~M_\odot$ for those MSP-WD systems with orbital periods of $2-9~\rm hours$. This implies that donor stars already develop He cores with masses of $0.16-0.17~M_\odot$, and their rich-H envelope has been stripped when orbital periods are longer than $2~\rm hours$. Otherwise, GR as a dominant mechanism will drive a new mass transfer stage and the binary system cannot experience a detached phase.

In the bottom panel of Figure 1, the evolution of donor-star radii is consistent with that in the top panel. It is clear that only two detached systems possess radii smaller than the Roche lobe radius of the NS binary with an orbital period of 9 hours. After the first mass transfer ceases, donor stars in those NS-MS binaries with initial orbital periods $\ge2.92~\rm days$  remain He cores and then evolve into WDs with masses of $0.16-0.22~M_\odot$. However, He WDs only in two detached MSP-WD systems with orbital periods shorter than 9 hours can fill the Roche lobe again due to orbital decay driven by GR and proceed to a second mass transfer in Hubble time. 

\begin{figure*}
\centering
\includegraphics[width=1.0\linewidth,trim={0 0 0 0},clip]{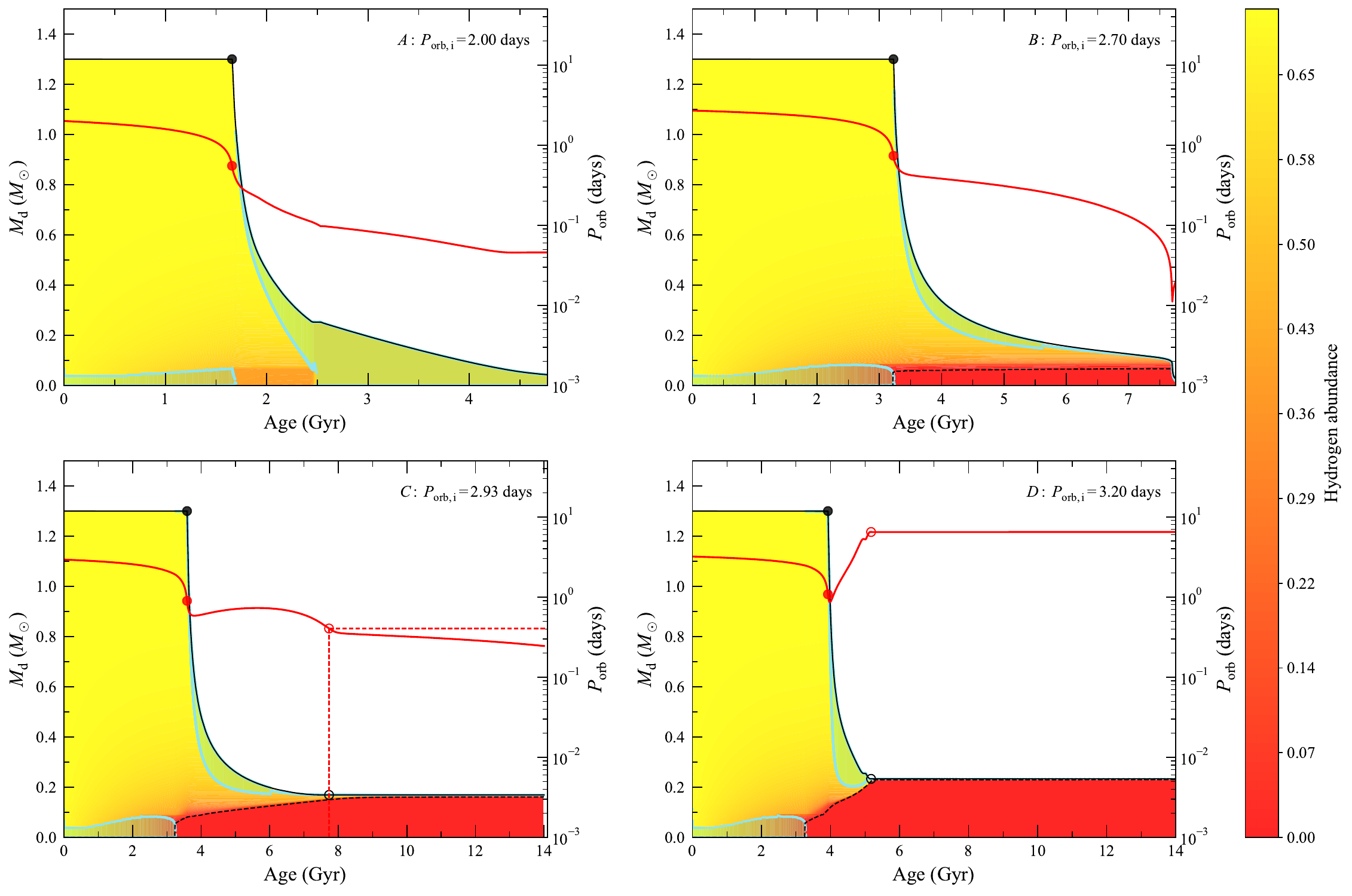}
\caption{Evolution of NS-MS binaries with $M_{\rm NS,i}=1.4~M_{\odot}$ and $M_{\rm d,i}=1.3~M_{\odot}$ in Kippenhahn diagram (left axis) and orbital period vs. stellar age (right axis) diagram. The solid and open circles represent the onset and end of the first mass transfer, respectively. The He core boundary is outermost location where H1 mass fraction is less than 0.01 and He4 mass fraction is greater than 0.1. The color bar indicates the hydrogen abundance at the corresponding mass coordinate of the donor star. The light green areas and the black dashed curves represent the convection regions and the helium core masses, respectively. The red curves correspond to the evolutionary tracks of orbital periods. Top-left, top-right, bottom-left, and bottom-right panels represent systems with initial orbital periods of $2.00~\rm days$, $2.70~\rm days$, $2.93~\rm days$, and $3.20~\rm days$, respectively.}
\label{s4}
\end{figure*}

The formation and evolution of He cores of donor stars play an important role in forming detached MSP-WD systems with short orbital periods. In Figure \ref{s4}, we show the Kippenhahn diagram of donor stars and the evolution of their orbital period for four systems in Figure \ref{s1}. Due to relatively long initial orbital periods, the systems B, C, and D have already developed He cores when mass transfer begins. Subsequently, the mass transfer will strip the H envelope of donor stars. It strongly depends on the orbital period when the H envelope is almost exhausted whether a system can evolve into a detached MSP-WD system with a short orbital period. In system B, the H envelope is exhausted until the orbital period decreases to $\sim0.01$ days due to slow mass transfer. However, GR already dominates mass transfer when the orbital period is $\sim0.1~\rm days$. Therefore, system B cannot evolve into a detached MSP-WD system. In system D, a rapid mass transfer causes its orbital period to increase to $\sim 6~\rm days$ when the donor star exhausts its H envelope and remains a He core. Only the orbital period of system C decreases to $\sim0.4~\rm days$ ($\sim9.6$ hours) when it evolves into a detached system consisting of a MSP and a He core. Subsequently, the orbital period decreases to a range of $2-9~\rm hours $ due to GR when the He core evolves into a He WD through a contraction and cooling process.

In system A, a short initial orbital period results in a short evolutionary timescale ($\sim1.6~\rm Gyr$) before RLOF. After mass transfer, the core of the donor star forms a He-rich zone with an H fraction of $\sim40\%$. The absence of a compact He core makes the helium-rich zone mix with the H-rich outer layers, resulting in significant chemical mixing throughout the whole donor star. Consequently, the donor star evolves to a fully convective state at $t\sim2.5~\rm Gyr$, leading to the cut off of the MB mechanism. Since the orbital period already decreases to shorter than 0.1 days, the subsequent mass transfer is dominated by GR, which drives the system to appear as a semidetached system rather than a detached system.

\begin{figure*}
\centering
\includegraphics[width=1.20\linewidth,trim={0 0 0 0},clip]{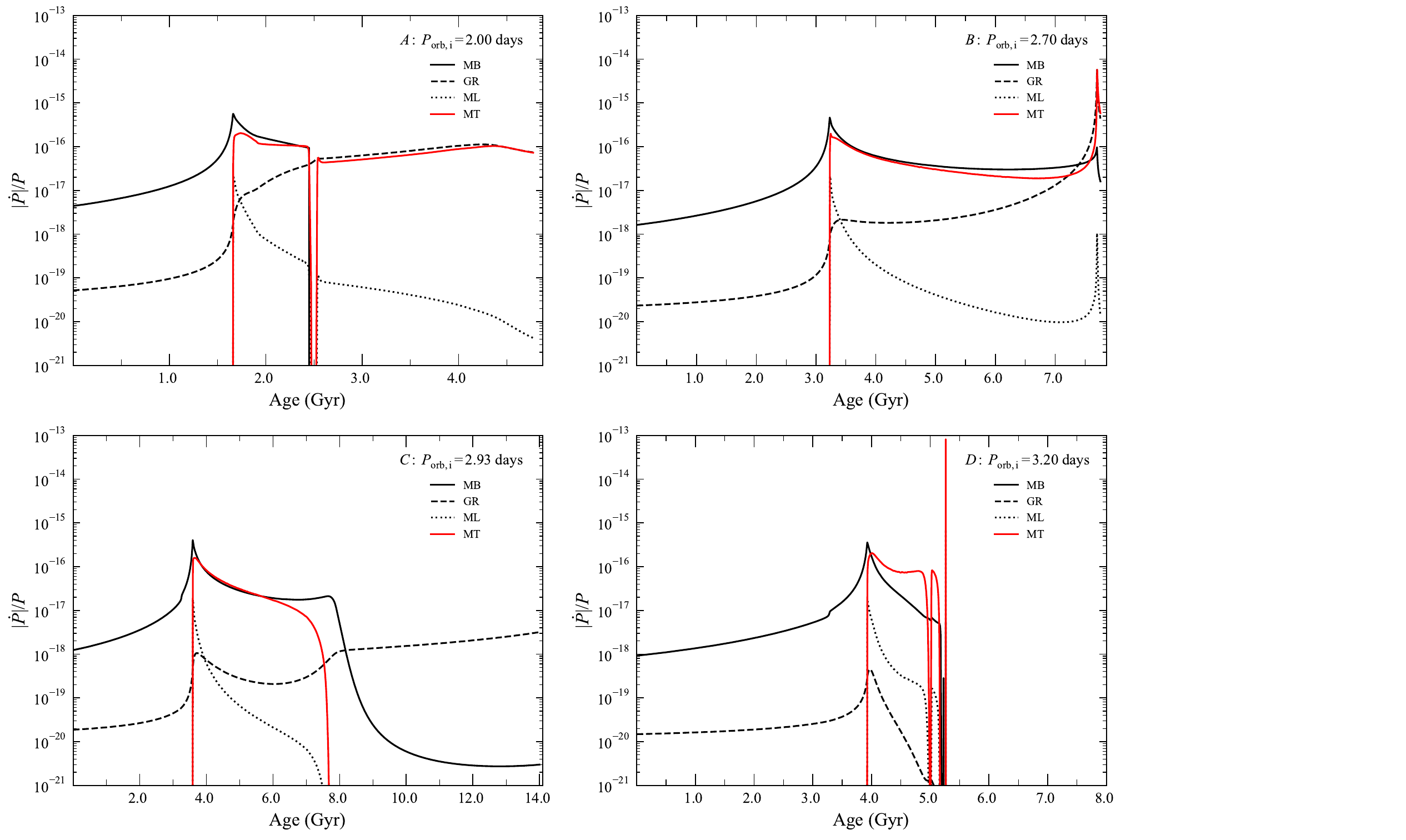}
\caption{Same as Figure.\ref{s4}, but for the evolution of four systems in $\dot{P}/P$ vs. stellar age diagram. The black solid, black dashed, black dashed-dotted, and red solid curves represent the orbital decay effect caused by standard MB, GR, mass loss (ML), and mass transfer (MT), respectively.}  \label{s5}
\end{figure*}

Figure \ref{s5} depicts the evolution of orbital decay effects caused by four different physical processes (MB, GR, mass loss, and mass transfer), in which $\dot{P}/P$ was derived from Equation (2). It is clear that orbital evolution depends mainly on a competition between the orbital decay driven by MB and the orbital expansion induced by mass transfer during RLOF. After the MB process of system A (the donor star becomes fully convective) stops at an age of $\sim2.5~\rm Gyr$ (corresponding to an orbital period of $\sim0.1$ days), GR drives the system to continue a mass transfer. A long evolutionary timescale develops a deep convective envelope in the donor star of system D, resulting in a high mass-transfer rate. Since mass transfer dominates orbital evolution, the orbital period of system D always increases after mass transfer begins. Therefore, these two systems cannot evolve into detached MSP-WD systems with short orbital periods. In system B, the strong MB process causes a rapid orbital decay and a shrinkage of the donor star. Because the radius and luminosity of a star uniquely depend on its degenerate core mass \citep{refs70}, the He core cannot grow.  As a consequence, the strong MB mechanism dominates mass transfer until an age of $\sim7.3~\rm Gyr$, after which the GR continues to drive mass transfer because of a short orbital period of $\sim0.1~\rm days$ (see also Figure 2).

System C can successfully evolve toward a detached MSP-WD system with a short orbital period. The progenitors of this kind of peculiar binary MSPs must satisfy the following properties: first, the donor star must develop a compact He core before RLOF; second, the He core can continuously grow, thus the orbital period should increase after mass transfer starts (a longer orbital period corresponds to a donor star with a larger radius, thus a heavier He core); third, the H envelope of the donor star is almost stripped when the orbital period decreases to a range of 2 to 9 hours.

\subsection{Formation of binary MSPs in the CARB MB prescription}
\begin{figure*}
    \gridline{
        \fig{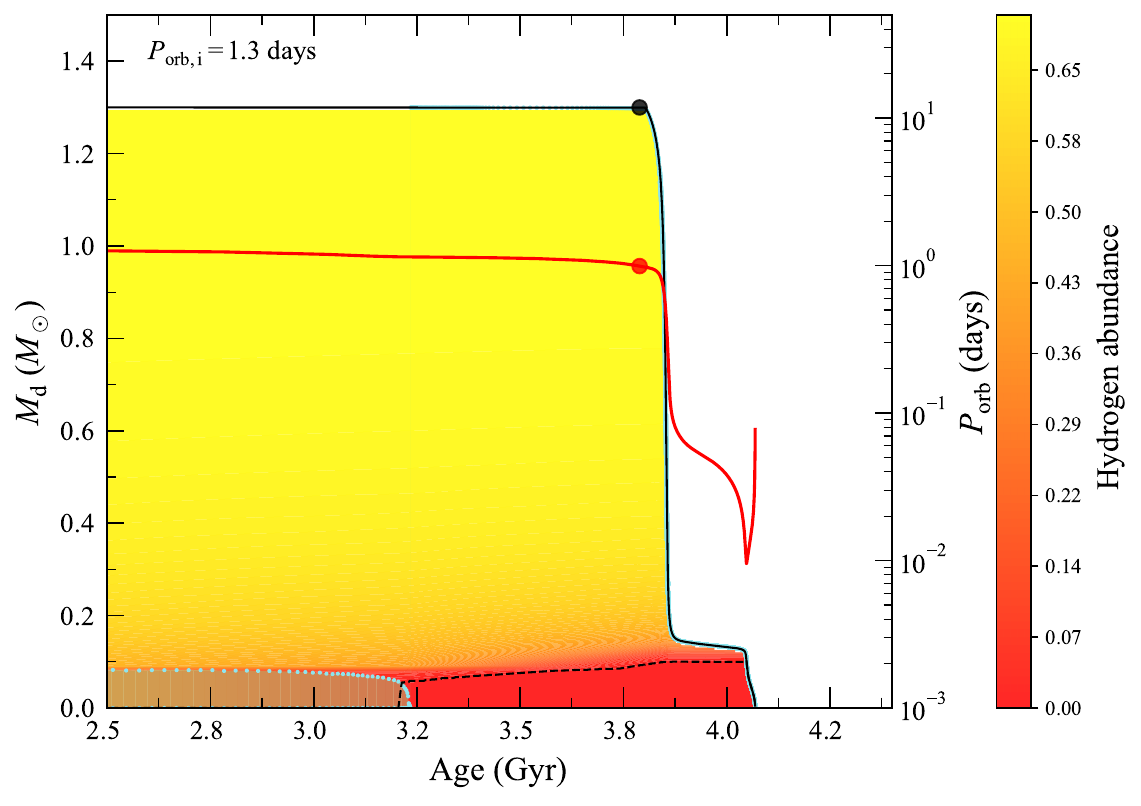}{0.45\textwidth}{}
        \fig{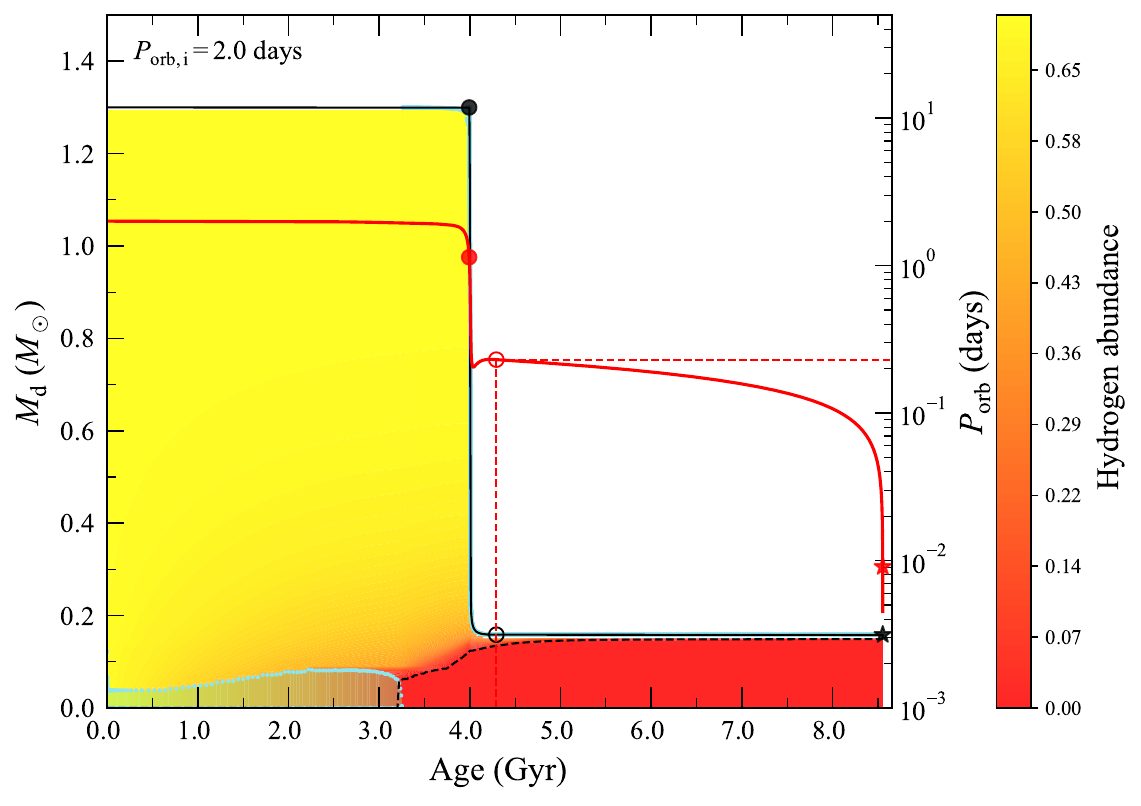}{0.45\textwidth}{}
    }
    \gridline{
        \fig{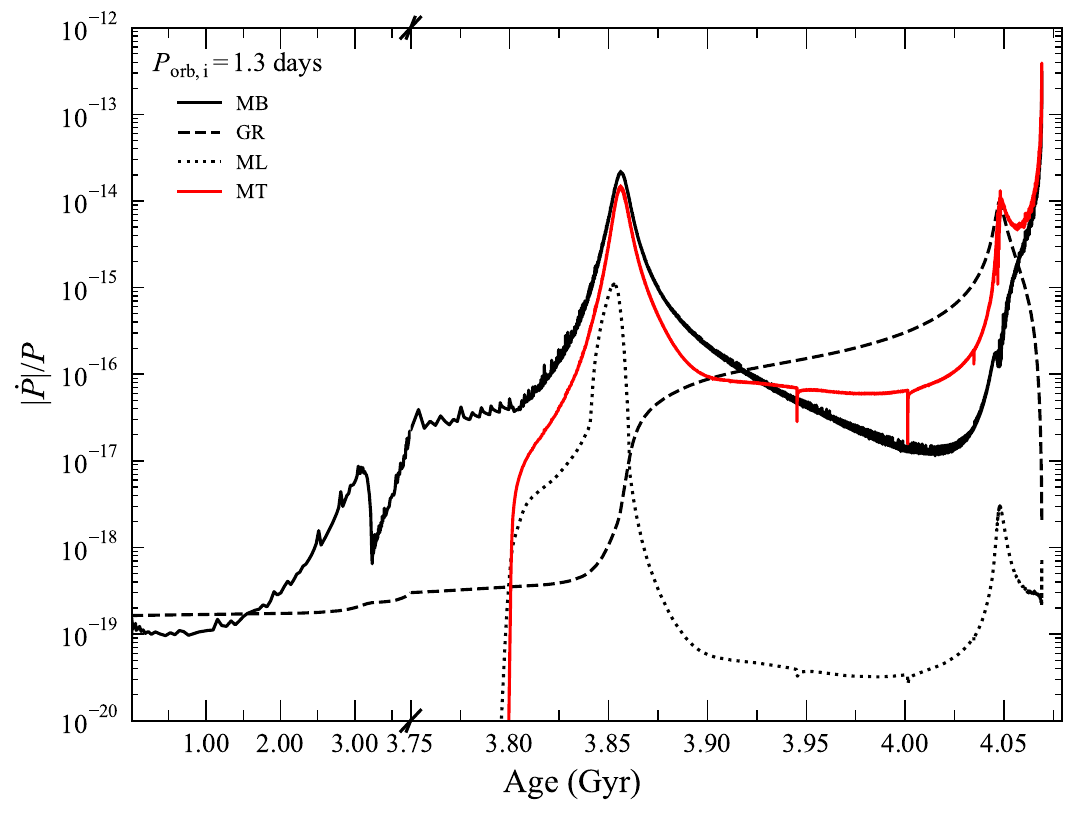}{0.45\textwidth}{}
        \fig{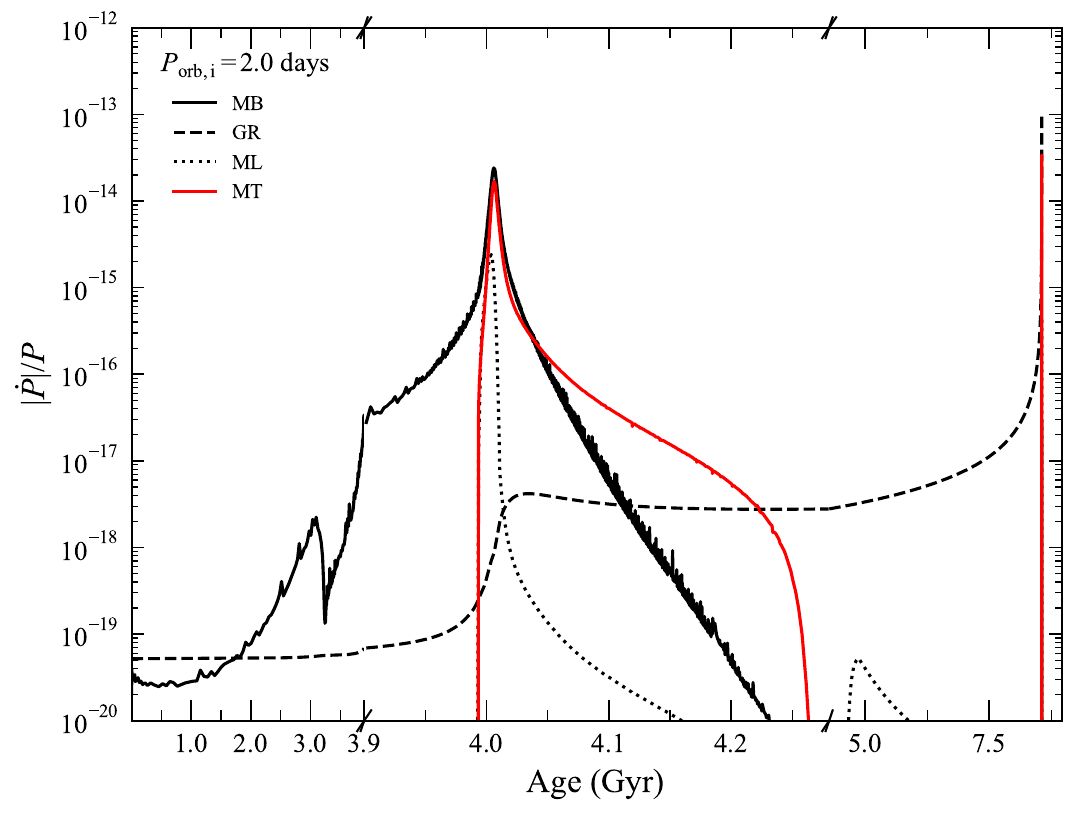}{0.45\textwidth}{}
    }
    \caption{Same as Figures~\ref{s4} and~\ref{s5}, but for NS-MS binaries with initial orbital periods of $P_{\rm orb,i} = 1.3$ (left panels) and $2.0~\mathrm{days}$ (right panels) in the CARB MB case.}
    \label{carb1}
\end{figure*}

Comparing with the standard MB law, the CARB MB law produces a relatively low rate of angular momentum loss for a star without filling Roche lobe, giving rise to a long evolutionary timescale before RLOF. In Figure \ref{carb1}, the donor star with $P_{\rm orb,i}=2.0~\rm days$ fills the Roche lobe at $t\sim4.0~\rm Gyr$, which is longer than that ($\sim1.6~\rm Gyr$) in the standard MB case. Due to a long timescale, the donor star already develops a He core with a mass of $0.122~M_\odot$ before RLOF. Subsequently, the strong MB mechanism leads to a high mass transfer rate, which causes a strong orbital expansion effect that exceeds the orbital decay effect due to angular momentum loss (see also Figure \ref{carb1}). During the orbital expansion, the He core continuously grows. As in $P_{\rm orb}\sim0.2~\rm days$, the most H-rich envelope is removed. When the mass transfer stops, the system evolves into a detached system consisting of an MSP and a He core. However, the MB mechanism dominates the orbital evolution of the system with $P_{\rm orb,i}=1.3~\rm days$ until the orbital period decreases to $<0.1~\rm days$. Subsequently, the GR becomes the dominant mechanism of orbital evolution and causes the orbit to continuously decay until a shortest orbital period. Therefore, this system always proceeds mass transfer and cannot evolve into a detached system.

\begin{figure}
\centering
\includegraphics[width=1.0\linewidth,trim={0 0 0 0},clip]{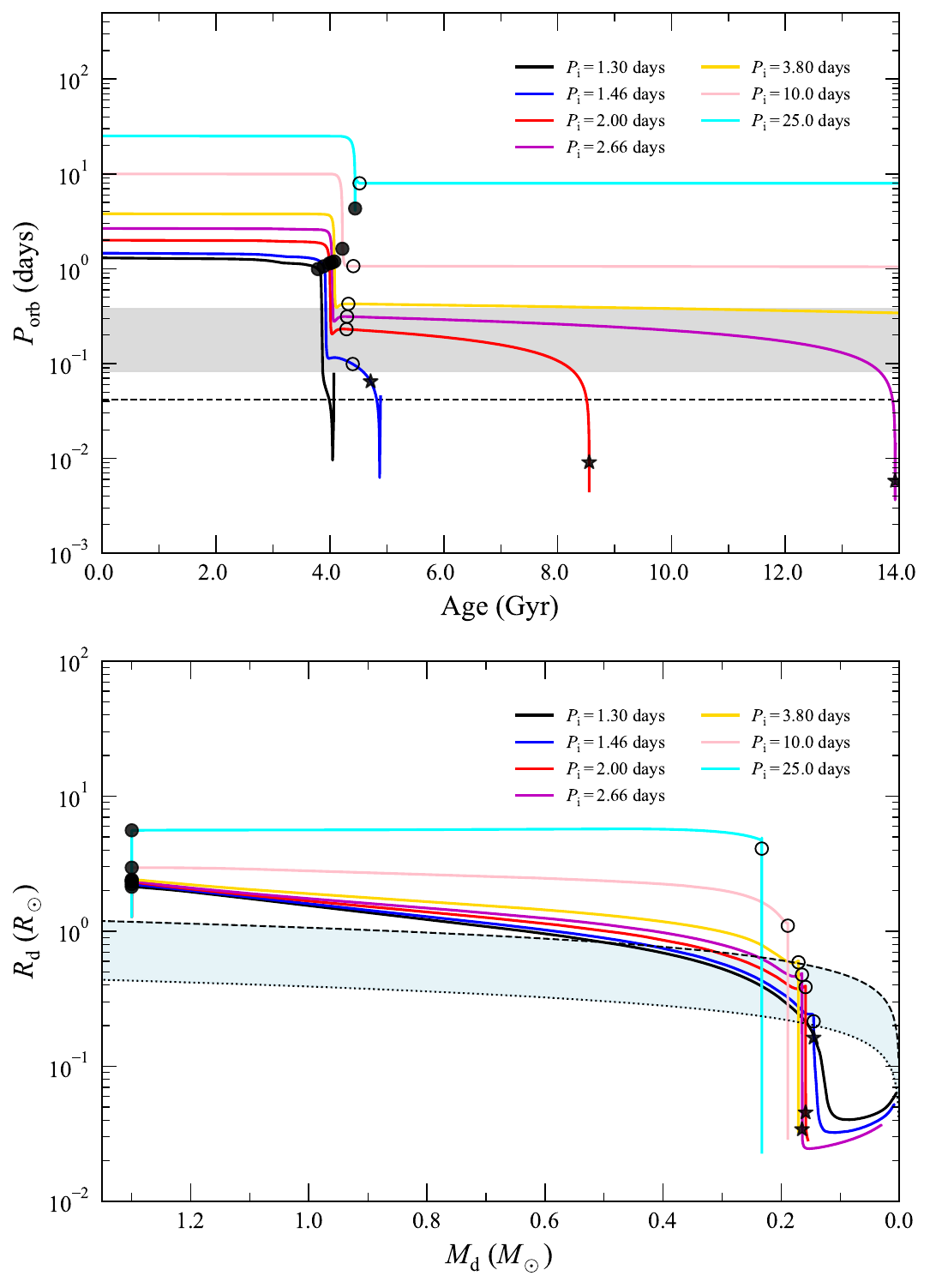}
\caption{Same as Figure \ref{s1}, but for $P_{\rm orb,i}=1.3-25~\rm days$ and the CARB MB law.}  \label{c1}
\end{figure}

Figure \ref{c1} summarizes the evolution of seven NS-MS binaries with $M_{\rm NS,i}=1.4~M_\odot$, $M_{\rm d,i}=1.3~M_\odot$, and $P_{\rm orb,i}=1.3-25~\rm days$ under the CARB MB case. It is clear that those NS-MS binaries with $P_{\rm orb,i}=1.46-2.66~\rm days$ can evolve into detached MSP-WD systems with short orbital periods of $2-9~\rm hours$. This orbital period range is much wider than that in the standard MB case, even if an anomalously large MB index of $\gamma=5$ was used \citep{istr14}. Adopting the intermediate MB prescription given by \cite{van19mn}, a detailed binary evolution model shows that NS-MS binaries with $M_{\rm NS,i}=1.3~M_\odot$, $M_{\rm d,i}=1.25~M_\odot$, and $P_{\rm orb,i}=7-15~\rm days$ can evolve towards detached MSP-WD systems with orbital periods of $2-9~\rm hours$ \citep{chen21}. Ignoring a tiny difference in initial masses, our initial orbital period range is narrower than that of \cite{chen21}, and moves toward a direction of short orbital period.

In particular, the CARB MB mechanism can produce MSP-WD systems with long orbital periods. Two systems with $P_{\rm orb,i}=10$ and $25~\rm days$ can evolve into detached MSP-WD systems with orbital periods of $\sim1$ and $8~\rm days$, respectively. However, the intermediate MB prescription can only produce detached MSP-WD systems with a longest orbital period of $\sim0.4~\rm days$ \citep[see also Figure 3 of][]{chen21}. Those NS-MS binaries with initial orbital periods longer than 25 days would experience dynamically unstable mass transfer, resulting in a common envelope phase \citep{chen21}. Therefore, the CARB MB law can help alleviate the fine-tuning problem of initial orbital periods in forming detached MSP-WD systems with short orbital periods. Meanwhile, it can also produce observed MSP-WD systems with long orbital periods, which cannot be formed in the intermediate MB scheme \citep{chen21}.

In the bottom panel of Figure \ref{c1}, one can see that a longer initial orbital period tends to remain a more massive He core when the first mass transfer stops. This is because a longer initial orbital period corresponds to a longer evolutionary timescale before RLOF, and donor stars can develop a more massive He core through a long-term nuclear evolution. As $P_{\rm orb,i}=25~\rm days$, the mass of the He core (pre-WD) is greater than $0.2~M_\odot$. The masses of other He cores are less than $0.2~M_\odot$. In those MSP-WD systems with orbital periods of $2-9~\rm hours$, the WD masses concentrate in a narrow range of $0.16-0.17~M_\odot$.

\begin{figure}[htbp]
    \centering
    \includegraphics[width=1.0\linewidth,trim={0 0 0 0},clip]{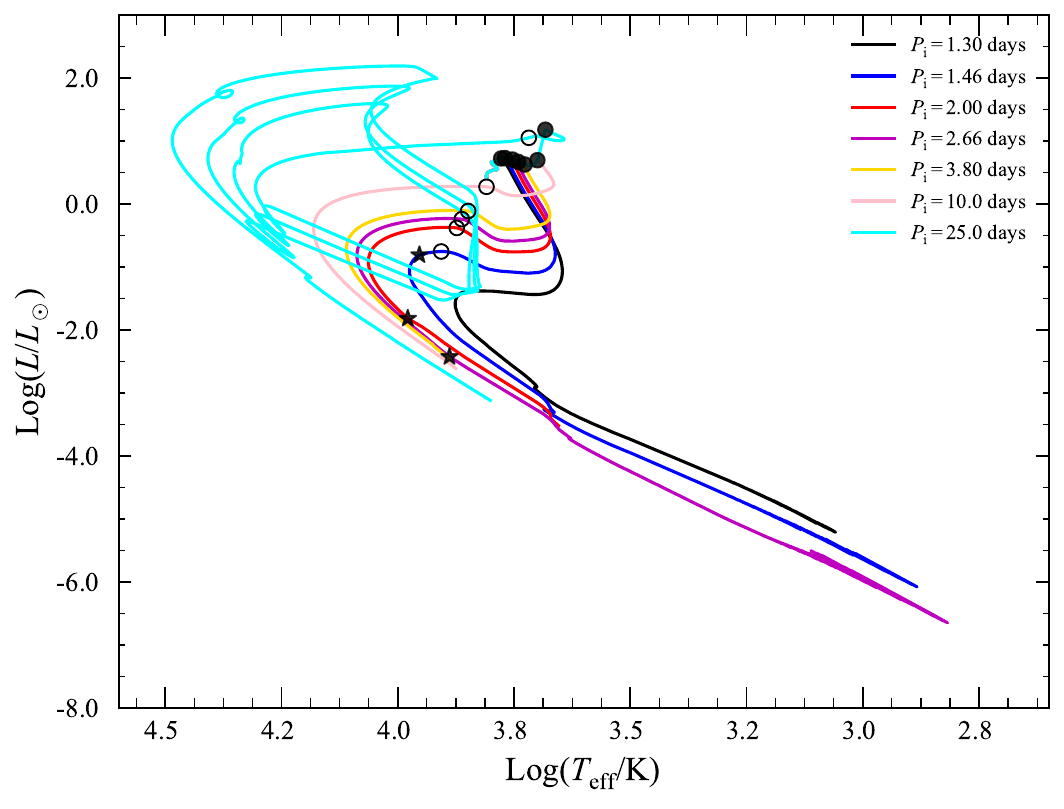}
    \vspace{0em} 
    \includegraphics[width=1.0\linewidth,trim={0 0 0 0},clip]{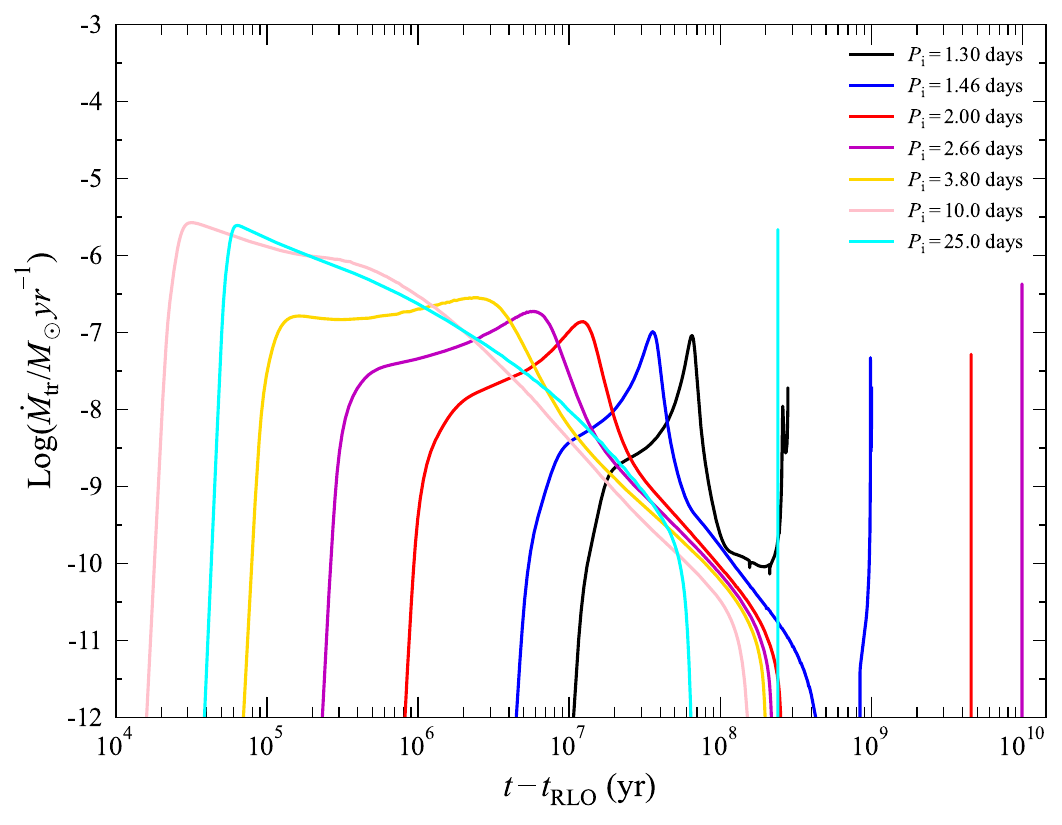}
    \caption{Evolution of NS-MS binaries in  Figure~\ref{c1} in Hertzsprung-Russell (HR) diagram (top panel) and the mass transfer rate vs. mass-transfer timescale ($t-t_{\rm RLO}$) diagram (bottom panel). }\label{hr}
\end{figure}

To better understand the evolutionary properties of donor stars and mass transfer, Figure \ref{hr} displays the evolution of the NS-MS binaries of Figure \ref{c1} in the HR diagram and the mass transfer rate versus the mass transfer timescale diagram. In the top panel, the donor star with $P_{\rm orb,i}=25~\rm days$ undergoes four hydrogen shell flashes. Our simulation indicate that this pre-WD mass is greater than $0.2~ M_\odot$, which is consistent with the pre-WD mass range ($0.212-0.319 M_\odot$) that experienced hydrogen shell flashes in a basic model with $Z = 0.02$ \cite[see also their table 2 of][]{istr16}. Our simulated other pre-WDs masses are less than $0.2~ M_\odot$, thus there are no hydrogen shell flashes during their evolution. When those donor stars fill Roche lobes again, the luminosities and effective temperatures are $\sim0.01-0.1~L_\odot$ and $\sim8000-10000~\rm K$, respectively, implying that they have already evolved into WDs.

The bottom panel of Figure \ref{hr} depicts the evolution of mass-transfer rates of NS-MS binaries in Figure \ref{c1}. Donor stars in two systems with long orbital periods of 10.0 and 25.0 days develop a deep convective envelope and transfer mass onto NSs at a relatively high rate of $10^{-7}-10^{-6}~M_\odot\,\rm yr^{-1}$. After the mass transfer of $78$ and $193~\rm Myr$, these two systems evolve into binary MSPs with long orbital periods. In the system with $P_{\rm orb,i}=25~\rm days$, the hydrogen-shell flashes will trigger a short-lived ($\sim 10^3~\rm yr$) episode of
mass transfer, in which the mass-transfer rate exceeds the Eddington accretion of the NS. After the first mass transfer of $\sim200-400~\rm Myr$, three systems with $P_{\rm orb,i}=1.46, 2.0$, and 2.66 days evolve into detached MSP-WD systems. Subsequently, WDs fill the Roche lobes to begin the second mass transfer due to orbital decay induced by GR and the systems appear as UCXBs. The system with $P_{\rm orb,i}=1.3~\rm days$ always goes through a mass transfer on a timescale of $\sim250~\rm Myr$ and cannot become detached a MSP-WD system.

\subsection{Initial parameter space of binary MSPs with He WDs}

\begin{figure*}
\centering
\includegraphics[width=1.0\linewidth,trim={0 0 0 0},clip]{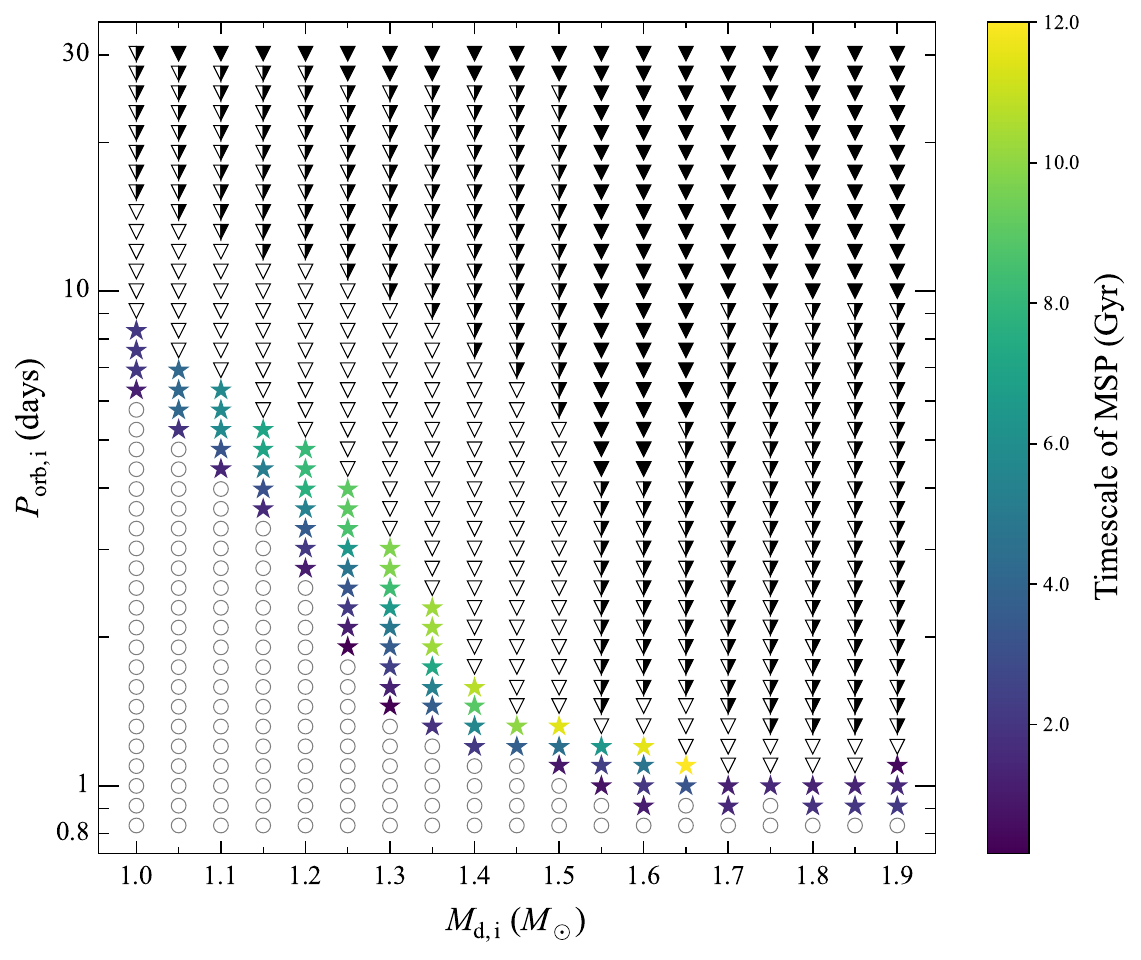}
\caption{Parameter space distribution of NS-MS binaries evolving into MSP-WD systems in the initial orbital period vs. initial donor-star mass diagram in the CARB MB case. The stars with different colors, open triangles, and semi-solid triangles, and solid triangles represent NS-MS binaries that can evolve into detached MSP-WD systems with orbital periods of $2-9~\rm hours$, $9~\rm hours-1~\rm day$, $1-10~\rm days$, and $>10~\rm days$, respectively. The initial masses of NSs were taken to be $1.4~M_\odot$. The color bar describes the timescale that binary systems appear as detached MSP-WD systems with short orbital periods.}
 \label{space}
\end{figure*}

To investigate the progenitor properties of binary MSPs with He WDs, we have modeled the evolution of a great number of NS-MS binaries under the CARB MB prescription. Figure \ref{space} summarizes the progenitor distribution of binary MSPs with different orbital periods in the $P_{\rm orb,i}-M_{\rm d,i}$ diagram. Those NS-MS systems with $M_{\rm d,i}=1.0-1.9~M_\odot$ and $P_{\rm orb,i}=0.9-8.5~\rm days$ can potentially evolve toward MSP-He WD systems with short orbital periods of $2-9~\rm hours$. The upper boundary is very close to the bifurcation periods, over which NS-MS systems cannot evolve into UCXBs in Hubble time \citep{sluy05a,sluy05b}. NS-MS binaries with initial orbital periods shorter than the bottom boundary always undergo a mass transfer and appear as semi-detached systems. Compared with the parameter space obtained by \cite{chen21}, our initial orbital period and donor-star mass ranges are narrower and wider than theirs, respectively. When initial periods are in the range of $0.9-1.2~\rm days$, the CARB MB mechanism can still drive some NS-MS binaries with $M_{\rm d,i}=1.6-1.9~M_{\odot}$ to evolve into compact MSP-WD systems after the donor stars create a convective envelope. A similar phenomenon was already noticed in investigating the evolution of black hole-MS binaries \citep{yang25}.

As $M_{\rm d,i}=1.4~M_\odot$, those NS-MS systems in an interval ($P_{\rm orb,i}=1.2-1.5~\rm days$) of initial orbital periods can evolve into MSP-He WD systems with short orbital periods. In the standard MB case, the initial orbital periods with $M_{\rm d,i}=1.4~M_\odot$ are concentrated in an extremely narrow interval of $3.39-3.43~\rm days$ even if an anomalously high MB index ($\gamma=5$) was adopted \citep{istr14}. Adopting the intermediate MB law given by \cite{van19mn}, those NS-MS binaries with $M_{\rm d,i}=1.4~M_\odot$ can evolve toward compact MSP-He WD systems in an initial orbital period range of $P_{\rm orb,i}\sim4-6~\rm days$ \citep{chen21}. However, the intermediate MB law cannot produce MSP-He WD systems with long orbital periods \citep{chen21}. Therefore, the CARB MB law can successfully help us alleviate the fine-tuning problem.

In Figure \ref{space}, we also present the distribution of initial donor-star masses and initial orbital periods of NS-MS binaries that can evolve into binary MSPs with orbital periods of $9~\rm hours-1~\rm day$, $1-10~\rm days$, and $>10~\rm days$. Most of NS-MS binaries with $M_{\rm d,i}\le1.5~M_\odot$ and $P_{\rm orb,i}\le10~\rm days$ will evolve toward binary MSPs with orbital periods shorter than 1 day. These initial parameter spaces of binary MSPs can be applied to the statistics study by population synthesis simulations in the future.

\section{Discussion}
\subsection{Relation between orbital periods and WD masses of binary MSPs}
Adopting the CARB MB prescription, the evolutionary products of NS-MS binaries in LMXB stage are detached MSP-WD systems in a wide initial orbital period range. Stellar evolution theory predicted a tight relation between the core mass of giants and their radius \citep{joss87}. During evolution, the giant in LMXBs always fills the Roche lobe, thus its radius is equal to the Roche lobe radius, which is related to the orbital separation and orbital period \citep{eggl83}. Once the giant envelope has almost been removed, its core will evolve into a WD through a contraction and cooling process. Therefore, the orbital period of the nascent MSP-WD system should be correlated with the WD mass \citep{savo87,rapp95,taur99}.

\begin{figure*}
\centering
\includegraphics[width=1.0\linewidth,trim={0 0 0 0},clip]{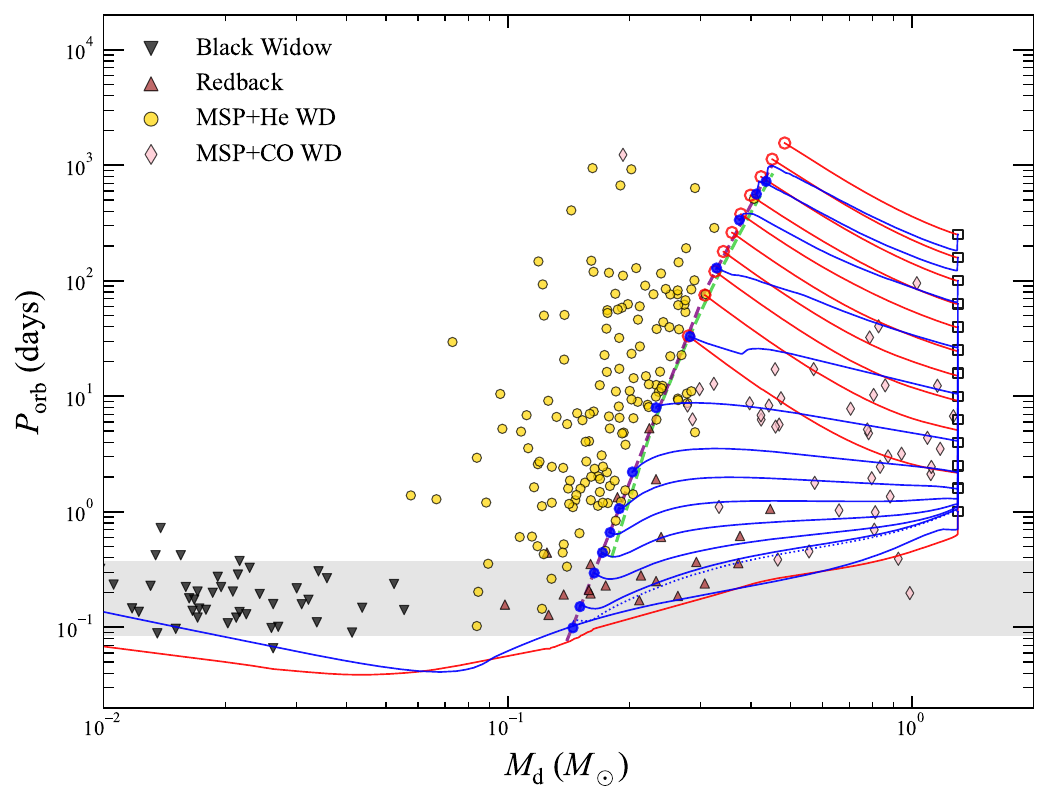}
\caption{Evolution of NS-MS binaries with $M_{\rm NS,i}=1.4~M_{\odot}$, $M_{\rm d,i}=1.3~M_{\odot}$, and different initial orbital periods in the orbital period vs. donor-star mass diagram. The red and blue curves correspond to the standard MB and the CARB MB cases, in which those red solid circles and black solid squares denote the positions when they evolve into detached MSP-WD systems. Circles with different colors represent the known binary MSPs (assuming an inclination angle $i = 90^{\circ}$ and $M_{\rm NS}=1.35M_{\odot}$). The horizontally shaded region corresponds to an orbital period range of $2-9~\rm hours$. The blue dotted curve represents the system which can evolve toward a detached MSP-WD system in a minimum initial orbital period ($P_{\rm orb,i} = 1.46~\rm days$). The data for binary MSPs originate from the ATNF pulsar catalog \citep[web version on August 29th, 2025,][]{manc05}. The green and purple dashed curve represent the relation between orbital period and WD mass of binary MSPs with He WDs obtained by \cite{taur99} and this work for Pop I stars.}
 \label{msp}
\end{figure*}

To investigate the relation between orbital periods and WD masses of binary MSPs, we model a grid of binary evolution of NS-MS binaries with $M_{\rm NS,i}=1.40~M_{\odot}$, $M_{\rm d,i}=1.30~M_{\odot}$, and ${\rm log}(P_{\rm orb,i}/\rm days)=0-2.4$ in steps of $\Delta{\rm log} (P_{\rm orb,i}/\rm days)=0.2$. Figure \ref{msp} summarizes the evolution of these NS-MS binaries in the orbital period vs. donor-star mass diagram. Due to the fine-tuning problem, those systems with ${\rm log}(P_{\rm orb,i}/\rm days)=0.6-2.4$ will evolve towards detached MSP-WD systems with orbital periods longer than $\sim40~\rm days$, but the system with ${\rm log}(P_{\rm orb,i}/\rm days)=0.4$ always appears as a semi-detached system with a mass transfer in the standard MB case. Therefore, an orbital period gap between 2 hours and $\sim40~\rm days$ is inevitable for detached MSP-WD systems.

The CARB MB mechanism can produce binary MSPs with a wide orbital period range of $2~\rm hours-800~\rm days$, which are in good agreement with the known binary MSPs with He WDs. In Figure \ref{msp}, most known binary MSPs with He WDs are located in the left of our simulated detached MSP-WD systems because their minimum WD masses were derived according to an inclination angle of $i=90^{\circ}$. The shortest initial orbital period that can form detached MSP-WD systems is 1.46 days, below which those systems will always appear as semi-detached binaries (these systems may be responsible for the formation of black widows). It should be noted that the NS-MS binary with 1.46 days can evolve into a detached MSP-WD system with a shortest orbital period of $\sim2~\rm hours$, which is consistent with the shortest orbital period observed in binary MSPs with He WDs.

Based on our simulated binary MSPs with He WDs, we fit a relation between the orbital period and WD mass (see also the purple dashed curve). In a range of WD masses of $\sim0.2-0.3~M_\odot$, our relation and the relation (the green dashed curve) given by \cite{taur99} overlap slightly. For the same WD mass, our simulated orbital period is slightly longer than that predicted by \cite{taur99} in the remaining range of WD masses. Different mass transfer schemes may be responsible for this discrepancy. \cite{taur99} calculated the mass transfer rate from $\dot{M}_{\rm tr}=PS[{\rm ln}(R_{\rm d}/R_{\rm RL})]^3\times10^3~M_\odot\,\rm yr^{-1}$ (where $PS(x)=[x+{\rm abs}(x)]/2$ and $R_{\rm RL}$ is the Roche-lobe radius of the donor star), however, we take the "Ritter" mass transfer scheme. Similarly, \cite{zhan21} found that mass-period relation from the models with the 'Kolb' mass transfer scheme is above that from the models with the 'classical' scheme.

\subsection{Formation of persistent UCXBs with relatively long orbital periods}

In the CARB MB case, our detailed binary evolution models show that NS-MS binaries can evolve into MSP-WD systems in a wide initial period range. Hydrodynamic simulations for those NS-WD binaries found that only binaries containing He WDs with masses less than $0.2~M_\odot$ can evolve toward UCXBs through a stable mass transfer \citep{bobr17}.  At present, there exist 20 identified UCXBs, which include 11 persistent sources and 9 transient sources \citep{arma23}. In globular clusters, several dynamic processes, such as direct collisions, tidal captures, and exchange interactions, could be responsible for the formation of some UCXBs. However, those dynamical processes are insignificant in the Galactic field. In an isolated binary evolution channel, UCXBs could evolve from NS binaries with MS, WD, or He donor stars. However, it is hard to account for all known UCXBs for the current binary evolution theory.

\cite{hein13} found that NS-WD systems cannot produce the high average mass-transfer rates ($>10^{-10}~M_\odot\,\rm yr^{-1}$) in three persistent UCXBs with relatively long orbital periods (40–60 minutes). Meanwhile, these three persistent UCXBs are also difficult to be interpreted by the NS-He stars channel \citep[see also Figures 4 and 5 of][]{wang21}. \cite{chen16} proposed an alternative channel, in which NS-MS binaries with intermediate-mass donor stars could evolve toward three persistent UCXBs with relatively long orbital periods if the donor star possesses an anomalously strong magnetic field ($\sim1000\rm G$).

\begin{figure}[h]
\centering
\includegraphics[width=1.0\linewidth,trim={0 0 0 0},clip]{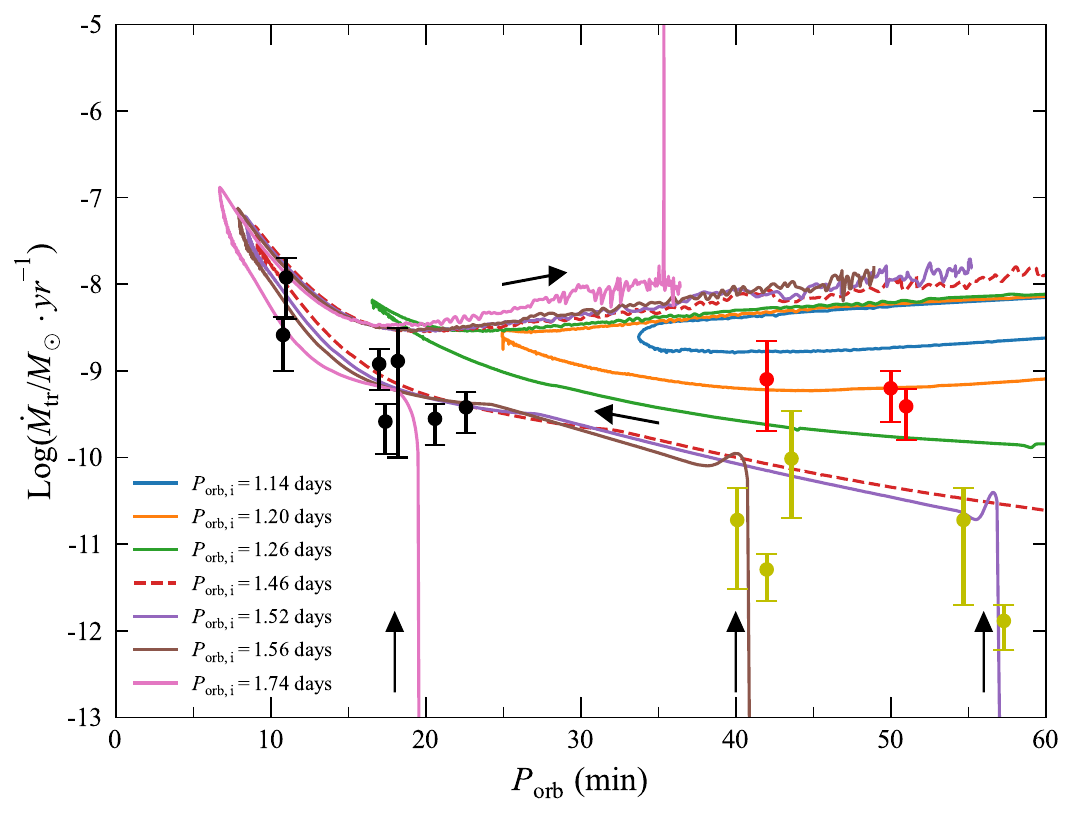}
\caption{Evolution of NS-MS binaries with $M_{\rm NS,i}=1.4~M_{\odot}$, $M_{\rm d,i}=1.3~M_{\odot}$, and different initial orbital periods in the mass-transfer rate vs. orbital period diagram under the CARB MB case. The black, red, and yellow solid circles with error bars denote the short-period persistent UCXBs, long-period persistent UCXBs, and transient UCXBs, respectively. The dashed curve represent the UCXB evolving from a detached MSP-WD system with a shortest orbital period whose progenitor is a NS-MS binary with an initial orbital period of $1.46~\rm days$ (see also Figure \ref{c1}).}
 \label{d2}
\end{figure}

Adopting the CARB MB law, NS-MS binaries can evolve into detached MSP-WD systems with short and long orbital periods, as well as semidetached accreting MSP-MS systems. To study whether the CARB MB prescription can produce these three persistent UCXBs with relatively long orbital periods, Figure \ref{d2} shows the evolution of several UCXBs evolved from NS-MS binaries in Figure \ref{c1}. It is noteworthy that the initial orbital period of 1.46 days (dashed curve) is a critical period, over which NS-MS binaries will evolve toward detached MSP-WD systems. It is clear that the evolutionary tracks of those detached MSP-WD systems are consistent with most persistent UCXBs with long periods and transient UCXBs. On the contrary, three persistent UCXBs with $P_{\rm orb}=40–50~\rm minutes$ could evolve from a NS-MS binary with $P_{\rm orb,i}=1.2~\rm days$, which always appears as a semi-detached system with mass transfer. \cite{seng17} also found that three persistent sources with relatively long orbital periods are approximately consistent with an UCXB evolved from a LMXB which always proceeds a mass transfer without experiencing a detached stage. However, their simulation can only account for the relatively high mass transfer of one source even if they take $\gamma=5$. Our simulations indicate that three persistent UCXBs with relatively long orbital periods are in the orbital decay stage, which, in principle, might be testable by long-term timing observation in the future.

\subsection{Detectability of detached MSP-WD systems as low-freuency GW sources}
After NS-MS binaries evolve into detached MSP-WD systems with short orbital periods, GR continuously drives their orbital decay, and some systems can emit low-frequency GW signals that can be detected by space GW detectors, such as LISA, TianQin, and Taiji. In the standard MB scheme, an NS-MS binary with $M_{\rm NS,i}=1.3~M_{\odot}$, $M_{\rm d,i}=1.4~M_{\odot}$, and $P_{\rm orb,i}\approx3.0~\rm days$ can successively evolve into a detached MSP-WD system with an orbital period of 4.8 hours and a UCXB \citep{taur18}. In the late stage of the binary MSP and in the early stage of the UCXB, the system is visible by LISA as a low-frequency GW source. However, the initial orbital period of those NS-MS systems needs to be fine-tuned.

\begin{figure}[h]
\centering
\includegraphics[width=1.0\linewidth,trim={0 0 0 0},clip]{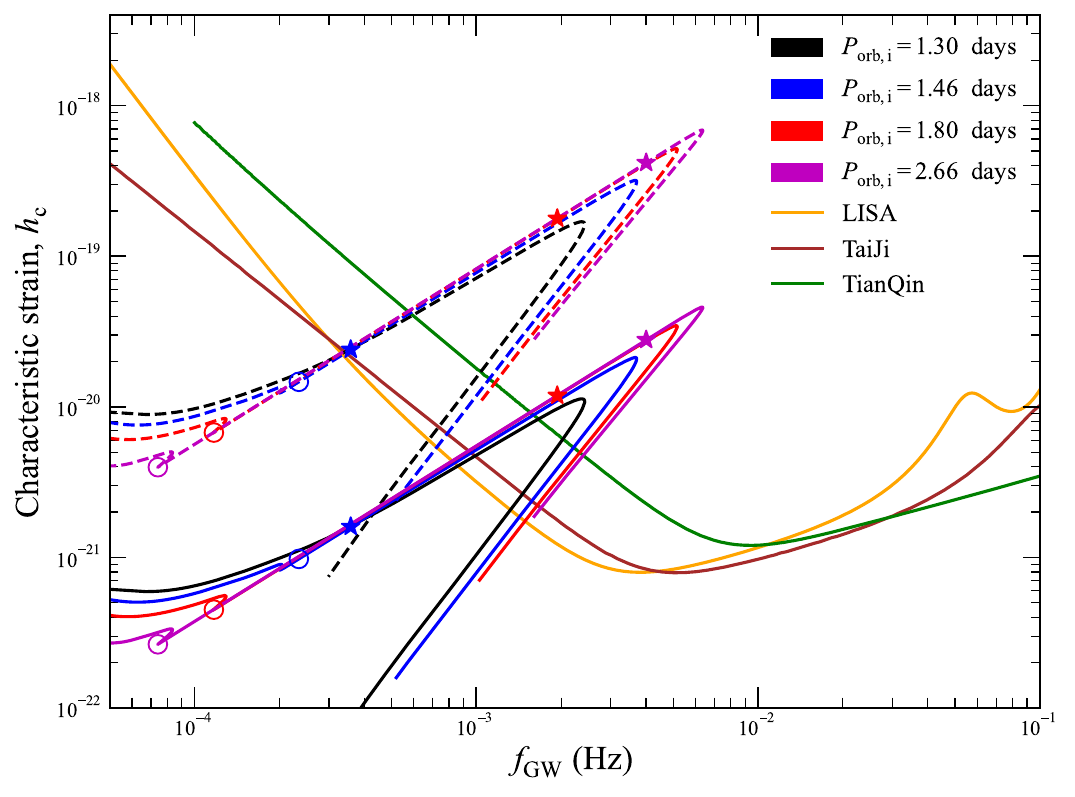}
\caption{Evolutionary tracks of four NS-MS binaries
with a $1.4~M_\odot$ NS, a $1.3~M_\odot$ MS companion star, and different initial orbital periods in the characteristic strain vs. GW frequency diagram. The yellow, purplish red, and green curves represent the sensitivity curves of LISA \citep{amar23}, Taiji \citep{ruan20}, and TianQin \citep{wang19}, respectively. The open circles and solid stars correspond to the end of first mass transfer and onset of second mass transfer. The distances of GW sources with solid curves and dashed curves are 15 kpc and 1 kpc, respectively.}
 \label{GW}
\end{figure}

Figure \ref{GW} shows the evolution of four NS-MS binaries which can evolve into compact detached MSP-WD systems and semi-detached systems in the characteristic strain versus GW frequency diagram. We use Equation (3) in \cite{chen20b} to estimate the characteristic strain of these sources in detached and semi-detached stages. When the distance is 1 kpc, three detectors can detect two systems with $P_{\rm orb,i}=1.8$ and 2.66 days in the late stage of detached MSP-WD systems. However, the shortest orbital period of the system with $P_{\rm orb,i}=1.46~\rm days$ is longer than that in the previous two systems, thus it cannot be detected in the detached stage. For a long distance of 15 kpc, the threshold frequencies that two sources with $P_{\rm orb,i}=1.8$ and 2.66 days are detectable as low-frequency GW sources are much higher than those at a distance of 1 kpc. For a low-mass WD without an H envelope, it fills the Roche lobe at an orbital period of $P_{\rm orb}\approx47.2~{\rm s}~(M_\odot/M_{\rm WD})=295~{\rm s}~(0.16~M_\odot/M_{\rm WD})$ \citep{chen20}. Therefore, a detached MSP-WD system with $0.16~M_\odot$ He WD should evolve into a UCXB when the GW frequency is $f_{\rm gw}=2/P_{\rm orb}=6.8~\rm mHz$. However, our detailed binary evolution models show that the critical GW frequencies are $\sim0.35-4.0$ mHz. This difference arises from different WD structures in which our simulated WDs have a thin H envelope. In the early stages of UCXBs, four sources can also be detected as low-frequency GW sources \citep{nele09}, depending on the distances, detectors, and initial orbital periods of the NS-MS binaries. 

\section{Summary}
In this work, we investigate the origin of fine-tuning problem about initial orbital periods in forming detached MSP-WD systems with short orbital periods of $2-9~\rm hours$. In the standard MB scheme, a high rate of angular momentum loss shortens the evolutionary timescale before RLOF, resulting in donor stars with short initial orbital periods not being able to develop compact He cores. However, a relatively low rate of angular momentum after RLOF produces a low mass transfer rate, which cannot strip most of the H envelope of donor stars to remain He cores when orbital periods decrease to shorter than $9~\rm hours$. Therefore, it is difficult for the standard MB mechanism to form detached MSP-WD systems with orbital periods of $2-9~\rm hours$. 

In the CARB MB case, NS-MS binaries with a $1.4~M_\odot$ NS, a $1.3~M_\odot$ MS donor star, and initial orbital periods of $P_{\rm orb,i}=1.46-2.66~\rm days$ can evolve into detached MSP-WD systems with short orbital periods of $2-9~\rm hours$. Meanwhile, those systems with initial orbital periods longer than 10 days can produce binary MSPs with long orbital periods of $>1~\rm days$. NS–MS binaries with an initial donor-star mass in the range of $1.0–1.9~M_\odot$ and an initial orbital period in the range of $0.9–8.5$ days could potentially evolve into detached MSP-WD systems with short orbital periods of $2-9~\rm hours$. Those systems over the boundary of the previous region could evolve toward binary MSPs with He WDs and long orbital periods. Our obtained initial parameter space of the progenitors of MSP-WD systems can be applied to future statistic study by population synthesis simulations. As a consequence, the CARB MB mechanism can alleviate the fine-tuning problem of initial orbital periods in forming compact MSP-WD systems discovered by \cite{istr14}, and can also produce MSP-WD systems with wide orbits which cannot be formed by the intermediate MB prescription \citep{chen21}.

\acknowledgments
We thank the anonymous referee for a very careful reading and
constructive comments that have led to the improvement of
the manuscript. This work was partly supported by the National Natural Science Foundation of China (under grant No. 12273014), and the Natural Science Foundation (under grant No. ZR2021MA013) of Shandong Province.

\bibliography{ref}
\bibliographystyle{aasjournal}
\end{document}